# The Effect of Na on Cu-K-In-Se Thin Film Growth

Christopher P. Muzzillo,[1,2] Ho Ming Tong,[2,3] Timothy J. Anderson[2]

[1]National Renewable Energy Laboratory, Golden, CO 80401, USA

[2]Department of Chemical Engineering, University of Florida, Gainesville, FL 32611, USA

[3]Oak Ridge National Laboratory, Oak Ridge, TN 37831, USA

Abstract

Co-evaporation of Cu-KF-In-Se was performed on substrates with varied surface Na compositions. Compositions of interest for photovoltaic absorbers were studied, with ratios of (K+Cu)/In ~ 0.85 and K/(K+Cu) ~ 0 - 0.57. Soda-lime glass (SLG) substrates led to the most Na by secondary ion mass spectrometry, while SLG/Mo and SLG/SiO$_2$/Mo substrates led to 3x and 3,000x less Na in the growing film, respectively. Increased Na content favored Cu$_{1-x}$K$_x$InSe$_2$ (CKIS) alloy formation by X-ray diffraction (XRD), while decreased Na favored the formation of CuInSe$_2$ + KInSe$_2$ mixed-phase films. Scanning electron microscopy and energy dispersive X-ray spectroscopy revealed the KInSe$_2$ precipitates to be readily recognizable planar crystals. Extrinsic KF addition also drove diffusion of Na out from the various substrates and into the growing film, in agreement with previous reports. Time-resolved photoluminescence showed enhanced minority carrier lifetimes for films with moderate K compositions (0.04 < K/(K+Cu) < 0.14) grown on Mo. Due to the relatively high detection limit of KInSe$_2$ by XRD and the low magnitude of chalcopyrite lattice shift for CKIS alloys with these compositions, it is unclear if the lifetime gains were associated with CKIS alloying, minor KInSe$_2$ content, or both. The identified Na-K interdependency can be used to engineer alkali metal bonding in Cu(In,Ga)(Se,S)$_2$ absorbers to optimize both initial and long-term photovoltaic power generation.

Highlights

- Increased substrate Na favored Cu$_{1-x}$K$_x$InSe$_2$ formation
- Decreased substrate Na favored the formation of CuInSe$_2$ + KInSe$_2$ mixed-phase films
- Extrinsic KF addition drove Na diffusion out from the substrate
- Moderate K compositions on Mo substrates had enhanced minority carrier lifetimes

1. Introduction

Recent reports have detailed power conversion efficiency enhancements when potassium fluoride and selenium have been co-evaporated onto Cu(In,Ga)Se$_2$ (CIGS) absorbers at around 350°C (KF post-deposition treatment (PDT)) [1-11]. Although the mechanism(s) responsible for these efficiency improvements are not clear, the KF PDT has been associated with multiple phenomena: *increased* hole concentration (e.g., by consuming In$_{Cu}$ compensating donors to produce K$_{Cu}$ neutral defects [12]) [5-8, 11-14],

*decreased* hole concentration (e.g., by consuming $Na_{Cu}$, which produces $In_{Cu}$ compensating donors, and may also lead to an increased Mo/CIGS barrier for current flow [8]) [1, 8, 10], grain boundary passivation [5, 15], general defect passivation [2, 3], Cu-depleting chemical reaction(s) resulting in better near-surface inversion [1, 8, 10, 16] or decreased valence band energy [14, 17, 18], morphology changes resulting in increased CdS nucleation sites [2, 10], general changes in CdS growth [16], formation of a passivating K-In-Ga-Se [10, 19] or K-In-Se [18] interfacial compound, and modified Cu-Ga-In interdiffusion [6, 13]. Additional, as-yet-unsubstantiated hypotheses have also been proposed: K could reduce the surface work function [20], interfacial K-Se compounds could cause beneficial effects [15], and detrimental $KInSe_2$ formation could occur [6]. A PDT without KF (just Se) has also been shown to significantly alter CIGS absorbers [21, 22], calling into question what 'control' absorber is appropriate for KF PDT comparison. High absorber Na and K composition has also been linked to drastically accelerated degradation in photovoltaic (PV) performance [23], which undermines the industrial relevance of initial performance gains achieved with high Na and K levels. Specifically, Na and K diffused into and degraded the ZnO layer after 100 h in damp heat and light [23], underscoring the importance of understanding alkali metal bonding in CIGS. In a departure from work utilizing KF precursors and PDTs, recent work used Cu, KF, In, and Se co-evaporation to form $Cu_{1-x}K_xInSe_2$ (CKIS) alloys with K/(K+Cu) composition, or x, varied from 0 to 1 [24, 25]. Increasing x in CKIS was found to monotonically decrease the chalcopyrite lattice size, increase the band gap, and increase the apparent carrier concentration. On the other hand, moderate K compositions ($0 < x < 0.30$) exhibited significantly longer minority carrier lifetimes [24, 25], relative to x ~ 0 and x ≥ 0.30. That work laid a foundation for experimentally identifying and ultimately engineering K bonding in chalcopyrite films.

While the mechanisms underlying KF PDT effects remain uncertain, it has been established that a relatively large amount of K is present at the p-n junction in the most efficient solar cells [1]. These advances have heightened interest in Cu-K-In-Se material with group I-poor (i.e. (K+Cu)/In ~ 0.85) and K-rich (x > 0.30) compositions. Processing techniques have presently been used to study the effect of Na on phase formation in Cu-K-In-Se with those compositions. Most CIGS films are grown with Ga/(Ga+In) ~ 0.2 – 0.3 and intentional gradients in cation composition [1, 26, 27]. However, Ga has presently been excluded to simplify data interpretation. A constant rate, single temperature process was also chosen to achieve uniformity in depth and avoid compositional gradients–as KF [6, 13] and Na [28, 29] have been shown to affect cation diffusion in CIGS. In order to isolate the effects of Na composition on Cu-K-In-Se growth, different substrates were used to alter the amount of Na on the substrate surface during growth. Bare soda-lime glass (SLG) led to the highest Na content, while SLG/Mo and SLG/SiO$_2$/Mo substrates established 3x and 3,000x less Na in absorbers (by secondary ion mass spectrometry (SIMS)), respectively.

2. Experimental

Co-evaporation of Cu, KF, In, and Se was performed on substrates of SLG and SLG/Mo ("Mo") at 500°C, as previously detailed [25]. The (K+Cu)/In composition was maintained near 0.85 for all films, while K/(K+Cu) was varied between 0 and 0.58. For the low Na substrate, an SiO$_2$ diffusion barrier (~25 nm) was sputtered onto SLG, followed by sputtering of 200 nm Mo ("SiO$_2$/Mo"). Deposited film compositions were measured with X-ray fluorescence (XRF) and SIMS. Interference between $^{39}$K$^{41}$K$^+$ and $^{80}$Se$^+$ ions was observed in SIMS on films with high K content (x ≥ 0.38). Symmetric X-ray diffraction (XRD) was performed with a Rigaku Ultima IV diffractometer to determine structure and assist in phase identification, as previously reported [25]. Standard diffraction patterns were calculated from published phase structures. The film morphology was observed using scanning electron microscopy (SEM), and local composition measurements were made with energy-dispersive X-ray spectroscopy (EDS) at a 13 kV accelerating voltage. Room temperature time-resolved photoluminescence (TRPL) was performed on absorber films with a 905 nm laser (1.37 eV) under low-injection conditions, and the response was detected with a near-infrared photomultiplier tube responsive to photons in the range 0.92 to 1.31 eV (details of the fiber optic system have been published elsewhere [30]).

3. Results

First, CuInSe$_2$ (x ~ 0) was grown simultaneously on Mo and SLG substrates at 500°C. The resulting films were very homogenous, with surfaces of triangular facets, and very similar grain size for each substrate (top, middle, and bottom of Fig. 1, respectively). On moving to Mo/CKIS with x ~ 0.38, substantial inhomogeneity arose (left of Fig. 2). The precipitation of large, planar crystals oriented perpendicular to the substrate caused this inhomogeneity. The precipitates appeared identical to previously observed KInSe$_2$ crystallites [25]. On the other hand, SLG/CKIS with x ~ 0.38 was very homogenous, and had a roughened surface (relative to x ~ 0; right of Fig. 2), possibly due to an increased nucleation rate. The Mo/CKIS film with x ~ 0.38 was further examined with SEM/EDS, and its overall composition was similar to that expected from in situ measurements (Table I). Spot EDS showed that the bulk homogenous film, the planar precipitates, and rounder surface precipitates had moderate, high, and low K/(K+Cu) compositions, respectively (Fig. 3 and Table I). The trends in Fig. 3 and Table I were representative of all the measured locations, despite subsurface X-ray emission and K/In peak overlap degrading spot EDS accuracy and precision, respectively.

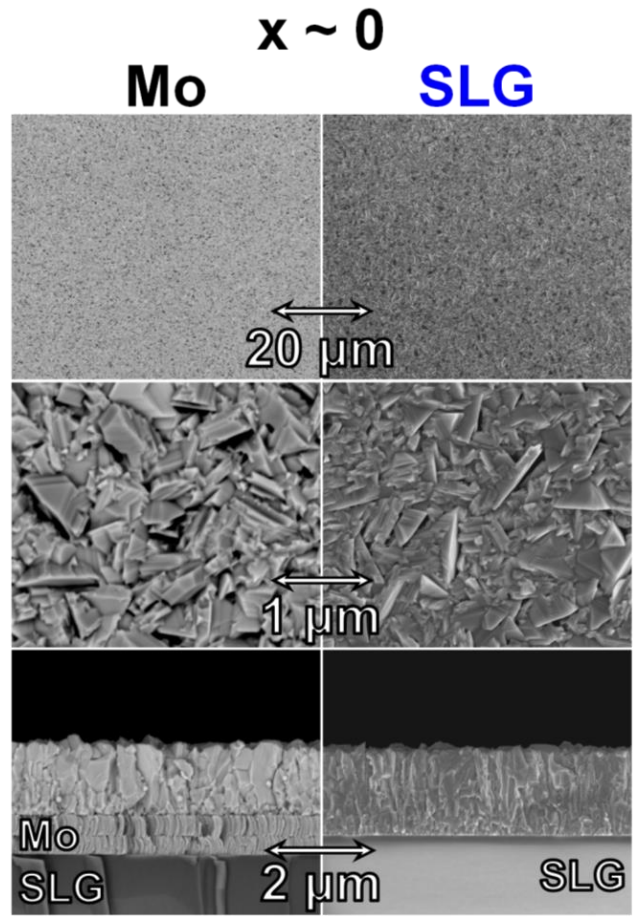

Fig. 1. Plan view (top) and cross-sectional (bottom) SEM micrographs of CuInSe$_2$ (CKIS with K/(K+Cu), or x ~ 0) films on Mo (left) and SLG (right) substrates.

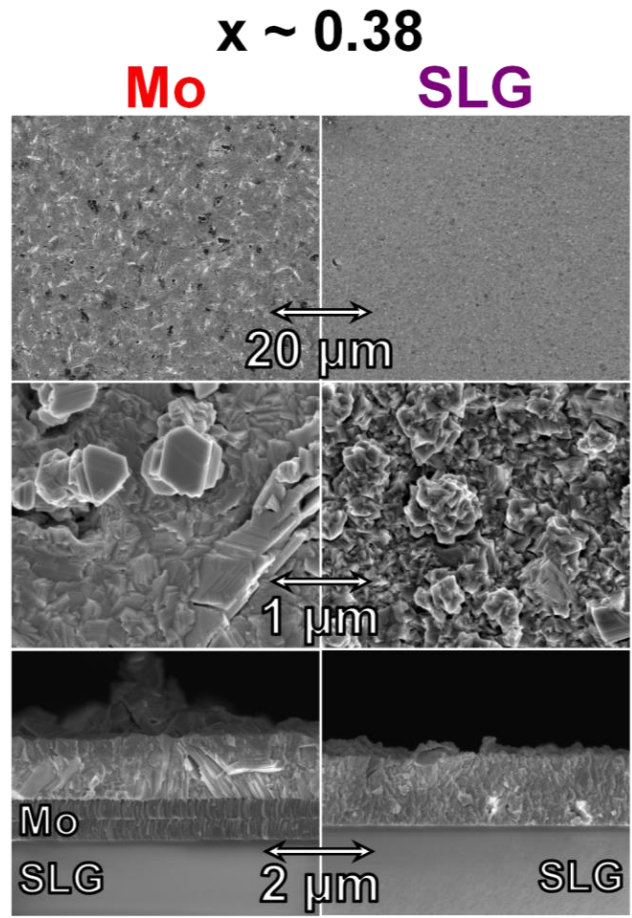

Fig. 2. Plan view (top) and cross-sectional (bottom) SEM micrographs of CKIS with K/(K+Cu), or x ~ 0.38 films on Mo (left) and SLG (right) substrates.

Table I. Atomic percent of each element and K/(K+Cu), or x, measured in situ, or by EDS on the overall film, spot a, spot b, and spot c (see Fig. 3). The substrate was Mo; in situ Se was added in excess, and assumed to be 50.0%.

| Measurement | Cu | K | In | Se | K/(K+Cu) |
|---|---|---|---|---|---|
| In situ | 13.8 | 8.6 | 27.6 | (50.0) | 0.376 |
| Overall EDS | 14.4 | 12.4 | 21.1 | 52.1 | 0.463 |
| EDS spot a | 16.1 ± 1.2 | 14.4 ± 3.3 | 23.9 ± 3.8 | 45.6 ± 1.3 | 0.468 ± 0.065 |
| EDS spot b | 14.4 ± 1.2 | 22.1 ± 7.2 | 21.0 ± 6.5 | 42.4 ± 1.7 | 0.595 ± 0.068 |
| EDS spot c | 17.7 ± 0.4 | 12.5 ± 5.3 | 18.4 ± 4.7 | 51.3 ± 2.0 | 0.398 ± 0.115 |

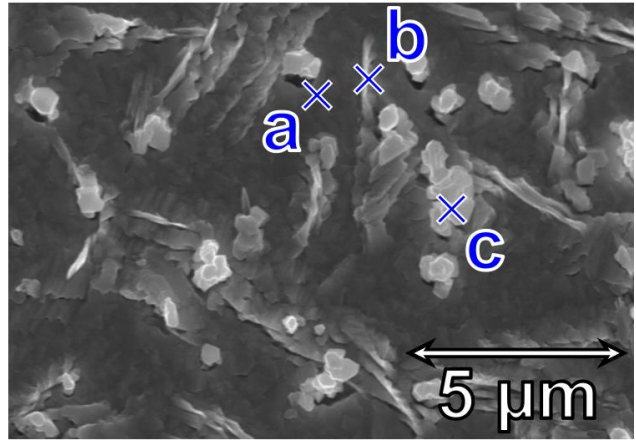

Fig. 3. Plan view SEM micrograph of CKIS with K/(K+Cu), or x ~ 0.38 film grown on Mo, where the labeled points correspond to EDS composition data in Table I.

The XRD scans of Mo/CuInSe$_2$ and SLG/CuInSe$_2$ were almost identical (Fig. 4). On moving to Mo/CKIS with x ~ 0.38, the only change was a decrease in the (101) plane's 17.1° 2θ peak. This peak also decreased for SLG/CKIS with x ~ 0.38, along with the (211) plane's 35.5° 2θ peak. Substantial shrinking of the chalcopyrite lattice was also observed for SLG/CKIS, and no other samples in Fig. 4. As expected, the peak shift was similar to that previously observed at x ~ 0.30 and 0.44 [25]. The apparent difference in extent of alloying (i.e. CKIS formation) between Mo and SLG substrates was repeatedly observed at high K compositions (x > 0.30). To demonstrate this, the extracted *a* lattice parameter was plotted against in situ K/(K+Cu) composition, or x, for various substrates in Fig. 5. As can be seen, the Mo and SiO$_2$/Mo substrates did not lead to significantly changed chalcopyrite lattices at high x composition. On the other hand, SLG substrates showed linear shrinking of the lattice at increased x, in line with the previous report [25]. As the chalcopyrite lattice can shrink from unintentional (K+Cu)/In reduction as well as intentional K/(K+Cu) increases, it remains unclear if CKIS alloys form at all on Na-poor substrates (Mo and SiO$_2$/Mo). The absence of Cu$_{2-x}$Se, K$_2$Se, CuIn$_2$Se$_{3.5}$, CuIn$_3$Se$_5$, CuIn$_5$Se$_8$, and K$_2$In$_{12}$Se$_{19}$ XRD peaks was an indication that substantial (K+Cu)/In deviation from set points did not occur. Transmission electron microscopy (TEM) will be needed to observe the relative extents of CKIS alloying and minor KInSe$_2$ formation in Na-poor conditions.

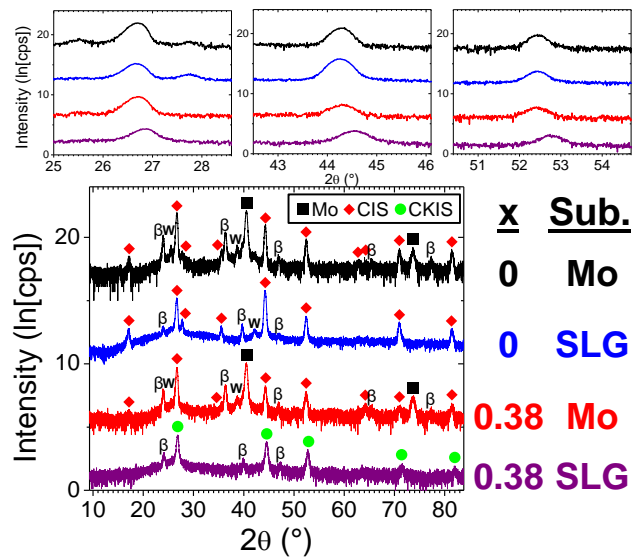

Fig. 4. Symmetric XRD scans from CKIS films with K/(K+Cu), or x ~ 0 on Mo (top; black) and SLG (2$^{nd}$; blue), and x ~ 0.38 on Mo (3$^{rd}$; red) and SLG (bottom; purple). Mo, CuInSe$_2$, and Cu$_{1-x}$K$_x$InSe$_2$ peaks are labeled with black squares, red diamonds, and green circles, respectively. 'β' and 'W' peaks are from Cu K$_β$ and W impurity in the Cu radiation source.

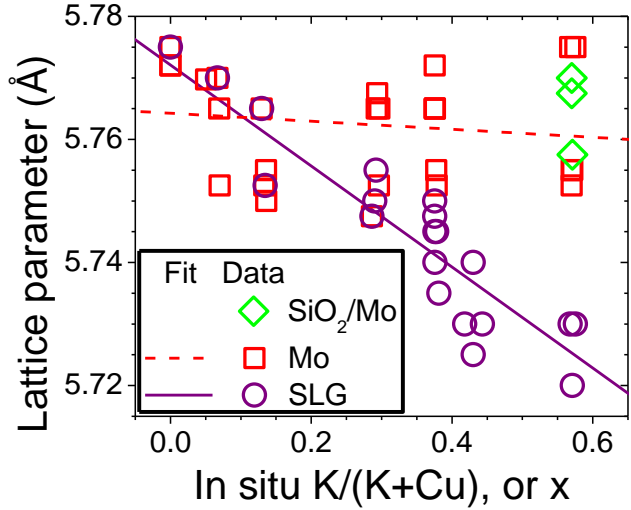

Fig. 5. CKIS chalcopyrite lattice parameter (*a*) versus in situ K/(K+Cu), or x composition for $SiO_2$/Mo (green diamonds), Mo (red squares), and SLG (purple circles) substrates. Lines are least squares fits to the Mo data (red dashed) and SLG data (purple solid).

In order to isolate the effect of substrate surface Na, CKIS with x ~ 0.57 was co-deposited onto $SiO_2$/Mo, Mo, and SLG substrates. As can be seen in Fig. 6, the $SiO_2$/Mo substrate led to a greater density of $KInSe_2$ precipitates than the Mo substrate, possibly due to an altered $KInSe_2$ nucleation rate. The chalcopyrite lattice peaks in XRD corresponded to $CuInSe_2$ (x ~ 0) for the $SiO_2$/Mo and Mo substrates, and a substantially smaller lattice for SLG (x ~ 0.57; Fig. 7), similar to the behavior in Fig. 5. XRD additionally revealed large amounts of $KInSe_2$ for the $SiO_2$/Mo substrate, significantly less $KInSe_2$ for the Mo substrate, and a very small amount of $KInSe_2$ for the SLG substrate. The monotonic trend in $KInSe_2$ formation was indirect evidence that the differences between the substrates were dominated by Na composition.

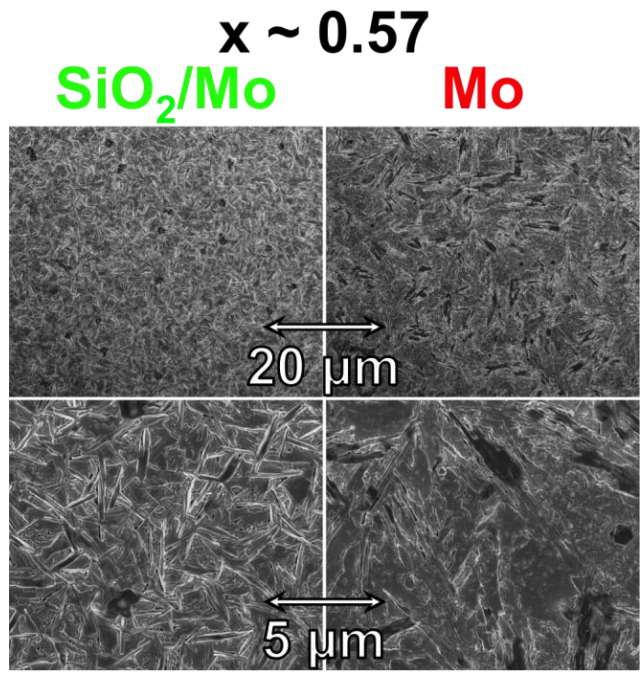

Fig. 6. High (top) and low (bottom) magnification plan view SEM micrographs of CKIS with K/(K+Cu), or x ~ 0.57 films on SiO$_2$/Mo (left) and Mo (right) substrates.

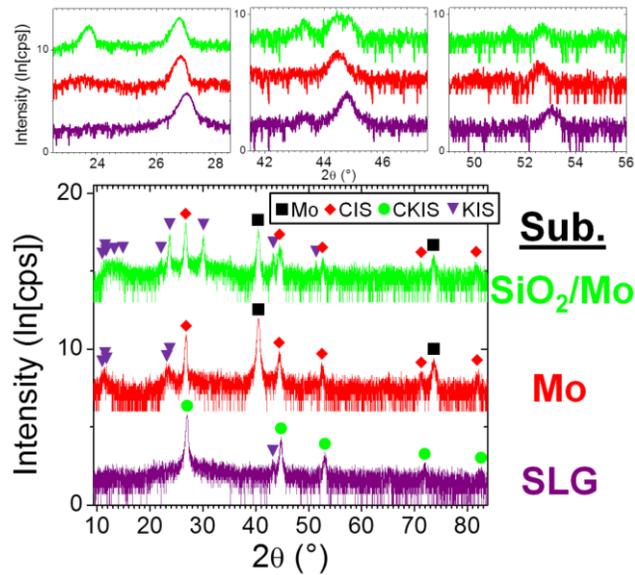

Fig. 7. Symmetric XRD scans from CKIS films with K/(K+Cu), or x ~ 0.57 on SiO$_2$/Mo (top; green), Mo (middle; red), and SLG (bottom; purple) substrates. Mo, CuInSe$_2$, Cu$_{1-}$

$_x$K$_x$InSe$_2$, and KInSe$_2$ peaks are labeled with black squares, red diamonds, green circles, and purple triangles, respectively.

To probe Na and K composition changes, SIMS was performed on x ~ 0 (CuInSe$_2$) films that were simultaneously deposited onto Mo and SLG substrates (Fig. 8). The Cu, In, and Se SIMS compositions were similar for all samples, and the profiles were relatively flat. The Na and K profiles were also mostly flat, with some instances of peaks at the free surface and substrate interface (Fig. 8), as expected from previous results [26, 27, 31]. The CuInSe$_2$ (x ~ 0) film grown on Mo exhibited SIMS Na and K signals about 3x and 5x lower, respectively, relative to material grown on SLG (Fig. 8). Similarly, the CKIS with x ~ 0.38 film had 2x less Na on Mo, relative to SLG. However, the CKIS x ~ 0.38 film exhibited effectively equal K compositions on Mo and SLG substrates. On changing from x ~ 0 to 0.38, the Na SIMS signal was 20x higher for both the Mo and SLG substrates, respectively. Finally, moving from x ~ 0 to 0.38 increased the K content 300x and 100x for the Mo and SLG substrates, respectively. For CuInSe$_2$ with x ~ 0, the SiO$_2$/Mo substrate led to 900x and 80x less Na and K than the Mo substrate, respectively (not shown). This translated to 3,000x and 400x less Na and K than the SLG substrate, respectively. As more Na than K diffused out from every substrate, it was assumed that substrate Na dominated the observed trends.

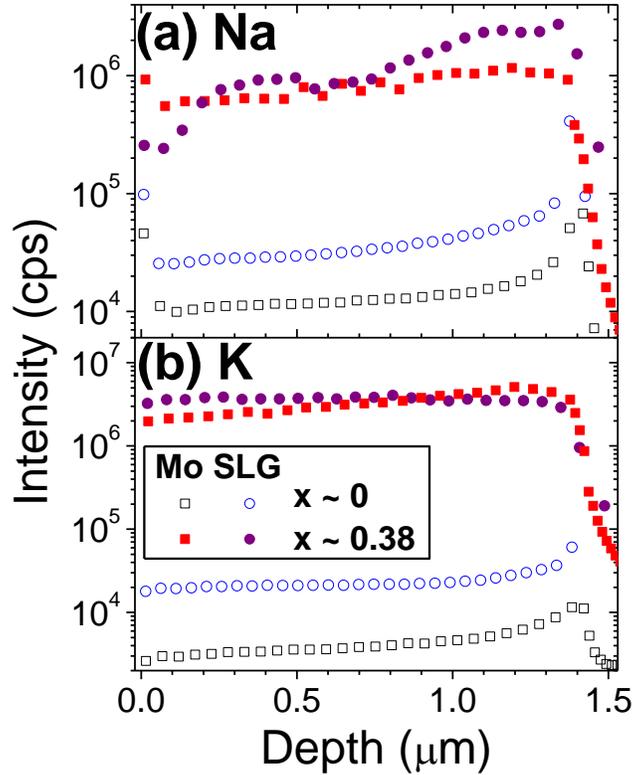

Fig. 8. SIMS Na (a) and K (b) profiles of CKIS with K/(K+Cu), or x ~ 0 on Mo (black empty squares) and SLG (blue empty circles), and x ~ 0.38 on Mo (red filled squares) and SLG (purple filled circles). A depth of 0 is the CKIS film's free surface; data is scaled so all CKIS films are equally thick.

Minority carrier lifetimes and majority carrier concentrations were estimated from TRPL measured on CKIS films soon after they were exposed to air, as previously outlined [25, 30]. The data in Fig. 9 were from samples grown with different processes, including substrate temperatures of 400 to 600°C. For Mo substrates, small-to-moderate K compositions (0.04 < x < 0.14) sometimes exhibited greatly enhanced lifetimes (Fig. 9 (a)), in agreement with the previous report [25]. This improvement in lifetime was never observed on SLG substrates at any composition. However, the x ~ 0 SLG substrates also had reduced lifetimes, relative to Mo substrates' best values. The carrier concentrations varied substantially, and showed no obvious trends on switching from Mo to SLG substrates at any composition.

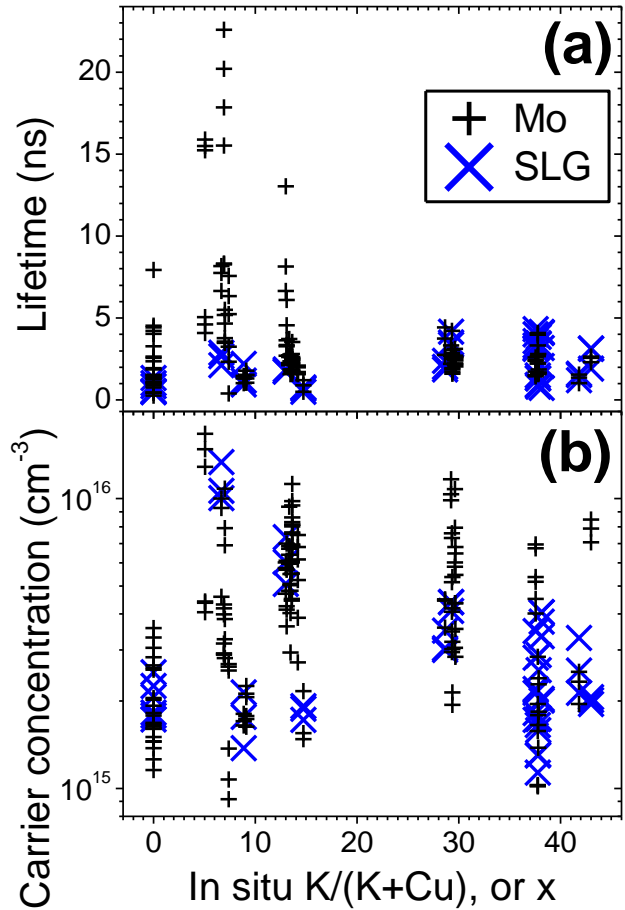

Fig. 9. Minority carrier lifetime (a) and majority carrier concentration (b) calculated from TRPL versus K/(K+Cu), or x composition of the CKIS on Mo (black plusses) and SLG (blue crosses) substrates.

4. Discussion

The SEM/EDS results showed clear evidence of phase segregation on substrates with decreased Na content. Precipitates of high K/(K+Cu) composition formed recognizable planar crystals, similar to KInSe$_2$ morphology [25]. The combined XRD results further demonstrated that decreasing substrate Na favored CuInSe$_2$ + KInSe$_2$ formation, and disfavored CKIS alloying. The XRD detection limit for KInSe$_2$ was relatively high: films exhibited large phase fractions of KInSe$_2$ by SEM (left of Fig. 2 and Fig. 3) which showed no evidence of KInSe$_2$ by XRD (Fig. 4).

The SEM/EDS and XRD results indicated that the relative extents of formation of CKIS and CuInSe$_2$ + KInSe$_2$ were strongly influenced by the substrate. Furthermore, results from SiO$_2$/Mo, Mo, and SLG showed that Na composition on the substrate's surface was correlated with the extent of the following net reaction:

$$CuInSe_2 + KInSe_2 \leftrightarrow 2\, Cu_{0.5}K_{0.5}InSe_2 \qquad (1)$$

Here, greater Na composition favored the forward reaction. During co-evaporation Na could affect the thermodynamics or kinetics of KF physisorption, adsorbate diffusion, desorption (re-evaporation), and chemisorption (crystal growth). The roughened surface in Fig. 2 (right) and increased density of KInSe$_2$ crystals in Fig. 6 (left) may be first indications that Na changed the relative nucleation rates of CKIS and KInSe$_2$, although more study is needed establish these connections. The SIMS results further indicated that the introduction of extrinsic KF acted as a driving force for diffusion of Na out from the substrate (Fig. 8 (a)). A previous report of CIGS grown on Na$_2$O- and K$_2$O-containing enamels found very similar SIMS Na and K profiles in the CIGS [13]. Another group used SIMS and inductively-coupled plasma mass spectrometry (ICPMS) to observe reduced Na composition after a KF PDT, rinse, and CdS chemical bath deposition (CBD) on alkali-free substrates [1]. They proposed that some chemical reaction favored K incorporation and Na removal [1]. A third report of a KF PDT on an alkali-free substrate found very similar SIMS Na and K profiles [5]. The overall Na composition was lower, but the surface Na composition was higher after the KF PDT and before a rinse. A fourth paper had decreased overall Na composition, with increased Na composition at the surface by SIMS and Auger electron spectroscopy (AES) when only the KF PDT was performed on an SLG/Mo substrate, without the rinse or CBD [7]. A fifth study again found reduced Na by SIMS after a KF PDT, rinse, and CBD on alkali-free substrates [8]. A sixth article showed similar Na composition in CIGS on alkali-free substrates by SIMS and ICPMS before and after a KF PDT, but before a rinse [10]. After the KF PDT and a rinse, however, the Na composition in the film dropped by an order of magnitude. Finally, increased Na levels from the substrate established similar SIMS profiles to K after a KF PDT and after Cu-KF-In-Ga-Se co-evaporation [11]. The above literature data are all in agreement that incorporation of K leads to the formation of some soluble Na-containing compound(s), as previously hypothesized [1]. The correlation between KF addition and diffusion of Na out from the substrates (Fig. 8) is therefore in excellent agreement with SIMS, AES, and ICPMS data from multiple laboratories [1, 5, 7, 8, 10, 11, 13]. However, it remains unclear if Na out-diffusion is driven by both chalcopyrite K-incorporation and $KInSe_2$ formation.

    The effect of Na on the sticking coefficient of KF is of interest, as any differences would stem from a change in the chemical potential of K during Cu-KF-In-Se co-evaporation. The SLG and Mo substrates led to similar K compositions by SIMS (Fig. 8 (b)). Moreover, it is unclear how the mixed-phase films change SIMS sputtering behavior. Therefore, the effect of substrate Na on the sticking coefficient of KF is uncertain. Previous work using a KF PDT found that substrate Na introduction did not

correlate with KF sticking coefficient after the PDT and ammonia rinse [10]. Another report found that Na content did not change KF sticking coefficient for a PDT before or after a subsequent ammonia rinse [11]. ICPMS will be needed to more accurately measure KF sticking coefficients and assess the effect of Na on the chemical potential of K.

The TRPL data showed that lifetimes for SLG substrates were consistently poor, even at $x \sim 0$. It is therefore unclear how the poor lifetimes associated with SLG substrates were affected by the CKIS alloying also observed on SLG. On the other hand, Cu-K-In-Se films with small x compositions ($0.04 < x < 0.14$) on Mo had reduced recombination rates very near the absorbers' surfaces, in agreement with the previous results [24, 25]. However, the present work has exposed possible $KInSe_2$ phase segregation in the films with superior lifetimes. The anticipated $KInSe_2$ phase fraction lies well below the detection limit of XRD. TEM could possibly enable direct observation of small $KInSe_2$ amounts, and their effect on lifetime. Furthermore, two-photon TRPL would help distinguish surface and bulk recombination rates in CKIS and $CuInSe_2$ + $KInSe_2$ films.

5. Conclusion

The effect of substrate surface Na on Cu-K-In-Se thin film growth was studied. Compositions of interest for PV absorbers were used (i.e. (K+Cu)/In ~ 0.85). Decreased Na content favored $CuInSe_2$ + $KInSe_2$ formation, while increased Na content favored CKIS alloying, as evidenced by SEM/EDS and XRD for $SiO_2$/Mo, Mo, and SLG substrates. SIMS showed that extrinsic KF addition drove Na diffusion out from the substrate, in agreement with previous work. Enhanced lifetimes were observed for moderate K compositions on Mo substrates (0.04 < K/(K+Cu) < 0.14), although it is unclear if these gains were associated with CKIS alloys, minor $KInSe_2$ impurity phase, or both. The identified Na-K interdependency is a valuable tool for studying alkali metal bonding in CIGS, and ultimately optimizing initial and long-term PV power conversion efficiency.

Acknowledgements

The authors thank Lorelle Mansfield, Carolyn Beall, Karen Bowers, and Stephen Glynn for assistance with experiments, Kannan Ramanathan and Stion Corporation for substrates, Clay DeHart for contact deposition, Matt Young for SIMS, and Bobby To for SEM. SEM/EDS was performed at the Center for Nanophase Materials Sciences at Oak Ridge National Laboratory. The work was supported by the U.S. Department of Energy under contract DE-AC36-08GO28308.